\documentclass[aps,prm,twocolumn,superscriptaddress,longbibliography]{revtex4-1} 
\usepackage[utf8]{inputenc}
\usepackage{bbm}
\usepackage{xcolor}
\usepackage{amsmath}
\usepackage{graphicx}
\usepackage{booktabs}
\usepackage[mathlines]{lineno}

\begin{document}

\title{Observation of giant circular dichroism induced by electronic chirality}

\author{Qian Xiao}\thanks{These authors contributed equally to this work.}
\affiliation{International Center for Quantum Materials, School of Physics, Peking University, Beijing 100871, China}

\author{Oleg Janson}\thanks{These authors contributed equally to this work.}
\affiliation{Institute for Theoretical Solid State Physics, IFW Dresden, Helmholtzstr. 20, 01069 Dresden, Germany} 

\author{Sonia Francoual}
\affiliation{Deutsches Elektronen-Synchrotron (DESY), Notkestr. 85, Hamburg D-22607, Germany}

\author{Qingzheng Qiu}
\affiliation{International Center for Quantum Materials, School of Physics, Peking University, Beijing 100871, China}

\author{Qizhi Li}
\affiliation{International Center for Quantum Materials, School of Physics, Peking University, Beijing 100871, China}

\author{Shilong Zhang}
\affiliation{International Center for Quantum Materials, School of Physics, Peking University, Beijing 100871, China}

\author{Wu Xie}
\affiliation{Deutsches Elektronen-Synchrotron (DESY), Notkestr. 85, Hamburg D-22607, Germany}
\affiliation{Center for Correlated Matte, Department of Physics, Zhejiang University, Hangzhou 310058, China}

\author{Pablo Bereciartua}
\affiliation{Deutsches Elektronen-Synchrotron (DESY), Notkestr. 85, Hamburg D-22607, Germany}

\author{Jeroen van den Brink}
\email{j.van.den.brink@ifw-dresden.de}
\affiliation{Institute for Theoretical Solid State Physics, IFW Dresden, Helmholtzstr. 20, 01069 Dresden, Germany}
\affiliation{W\"urzburg-Dresden Cluster of Excellence ct.qmat, TU Dresden, 01069 Dresden, Germany}
\affiliation{Institute for Theoretical Physics Amsterdam, University of Amsterdam, Science Park904, 1098 XH Amsterdam, The Netherlands}

\author{Jasper van Wezel}
\email{vanwezel@uva.nl}
\affiliation{Institute for Theoretical Physics Amsterdam, University of Amsterdam, Science Park904, 1098 XH Amsterdam, The Netherlands}

\author{Yingying Peng }
\email{yingying.peng@pku.edu.cn}
\affiliation{International Center for Quantum Materials, School of Physics, Peking University, Beijing 100871, China}
\affiliation{Collaborative Innovation Center of Quantum Matter, Beijing 100871, China}

\date{\today}

\begin{abstract}
Chiral phases of matter, characterized by a definite handedness, abound in nature, ranging from the crystal structure of quartz to spiraling spin states in helical magnets. In $1T$-TiSe$_2$ a source of chirality has been proposed that stands apart from these classical examples as it arises from combined electronic charge and quantum orbital fluctuations. This may allow its chirality to be accessed and manipulated without imposing either structural or magnetic handedness. However, direct bulk evidence that broken inversion symmetry and chirality are intrinsic to TiSe$_2$ remains elusive. Here, employing resonant elastic scattering of x-rays, we reveal the presence of giant circular dichroism up to $\sim$ 40$\%$ at forbidden Bragg peaks that emerge at the charge and orbital ordering transition. The dichroism varies dramatically with incident energy and azimuthal angle. Comparison to calculated scattering intensities unambiguously traces its origin to bulk chiral electronic order in ${\mathrm{TiSe}}_2$ and establishes resonant elastic x-ray scattering as a sensitive probe to electronic chirality.
\end{abstract}

\maketitle

Chirality can be found in various naturally occurring and artificial crystals such as elemental selenium and tellurium\,\cite{H.Fukutome}, $\alpha$-quartz\,\cite{chiralXRD_PRL,chiralXRD_PRB}, or berlinite\,\cite{chiralXRD_PRB}. The handedness of these materials allows for unique physical properties and applications\,\cite{CPGE}, but is hard-wired into their atomic structure and cannot easily be manipulated or switched. On the other hand, chiral or helical phases emerging in magnetic spin-density wave materials likewise break inversion and all mirror symmetries and endow their host materials with a handedness\,\cite{SDW}. In this case, however, controlling and employing the chirality requires large magnetic fields or spin currents. In this context, the novel type of chirality emerging from a weak-coupling charge and orbital instability proposed in $1T$-TiSe$_2$ stands apart: its origin in a charge density wave promises straightforward access, control, and switching by temperature, strain, or doping, all without breaking time-reversal or crystal symmetry.

The charge ordered and superconducting phases in the  quasi-two-dimensional van der Waals (VdW) material titanium diselenide can be easily accessed and tuned using external stimuli such as intercalation, pressure and gating\,\cite{CuxTiSe2,pressure_TiSe2,gating_TiSe2}. It is well-known to undergo a transition into a commensurate 2$\times$2$\times$2 charge density wave (CDW) state at $T_{\mathrm{CDW}}$ $\simeq$ 200\,K\,\cite{TiSe2_CDW}. After intense debate focusing on either a type of Jahn-Teller effect or exciton condensation underlying the CDW formation, a consensus has recently been reached that it is, in fact, stabilized by a combination of both\,\cite{EPC_2002prl,Exciton_JT_2002prb,excition_2007prl,combination_TiSe2}. Moreover, STM measurements suggested that the charge order in TiSe$_2$ may be chiral, spontaneously breaking spatial inversion symmetry and giving rise to domains of different handedness\,\cite{chiral_TiSe2_STM}.

Since STM is a surface-sensitive probe, it can only probe the periodic lattice distortion accompanying the CDW phase in the top atomic layer, precluding direct evidence for bulk inversion symmetry breaking. Indirect evidence in favor of a chiral CDW scenario has been found using other techniques\,\cite{chiral_TiSe2_183K,Iavarone}, but reports of the absence of such indirect evidence in alternative experiments have also persisted\,\cite{TiSe2_Seedge_prr,comment,STM_achiral,juan2023}, challenging the proposal of a chiral charge-ordered state in TiSe$_2$. Recently, the imprint of chirality in ${\mathrm{TiSe}}_2$ was discovered by the observation of polarization-dependent, bulk circular photogalvanic effect (CPGE) below 174\,K, but only when the compound is exposed to circularly polarized light as it cools into the CDW phase\,\cite{CPGE}. Circular light was suggested to favor one handedness during the growth of randomly distributed chiral domains, creating an observable net chirality.

\begin{figure}[htbp]
\centering
\includegraphics[width=\linewidth]{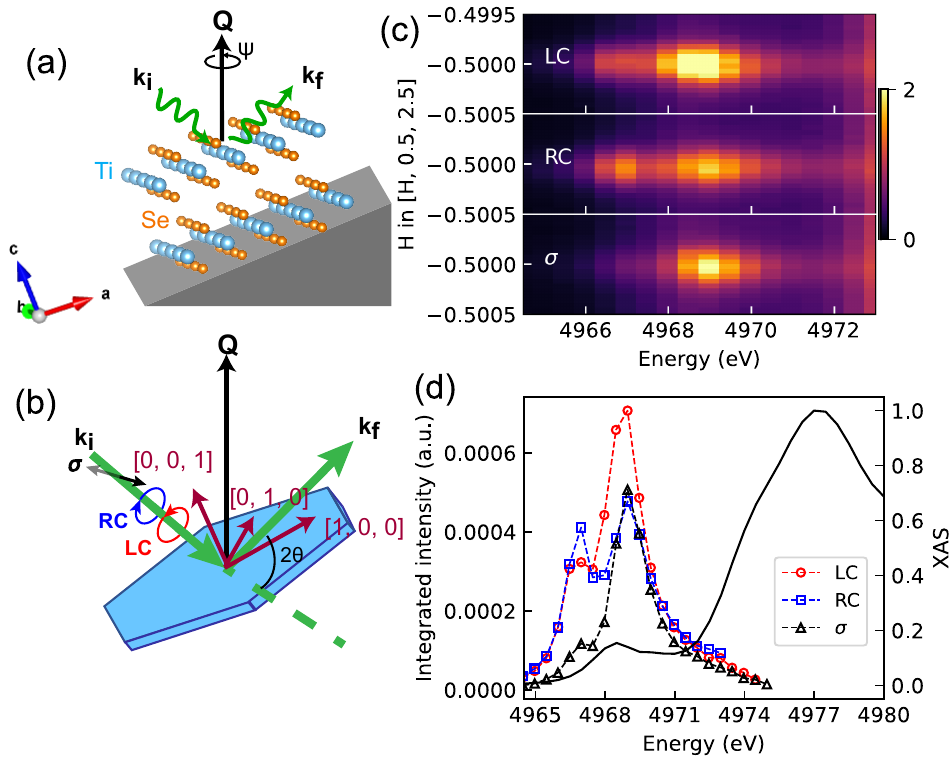}\caption{(a, b) Schematic plots of the experimental geometry (a) and polarizations (B) for the resonant elastic x-ray scattering (REXS) experiments, with a sketch of the crystal structure produced using VESTA\,\cite{VESTA}. The sample was glued on a wedge with an angle of 21.3$^\circ$. The azimuthal angle $\psi$ in (b) is 0$^\circ$ and the scattering plane is indicated by the light gray shaded area. (c) REXS intensity maps versus incident x-ray energy and momentum transfer [H, 0.5, 2.5] under left-circular (LC), right-circular (RC), and $\sigma$- polarized incident x-rays, respectively. Data were collected at a temperature of 6\,K. (d) Energy dependence of the momentum-integrated intensity under LC, RC, and $\sigma$-polarized incident x-rays. The integrated intensity was obtained by fitting the $\theta-2\theta$ scans collected at each incident photon energy shown in (c) with a Gaussian function and subtracting the fluorescence background (see Supplemental Material for details). All curves show the double resonance structure arising from the splitting of Ti $t_{2g}$ and $e_g$ orbitals. The black solid line shows the x-ray absorption spectrum (XAS) under $\sigma$-polarized incident x-rays.
\label{fig1}}
\end{figure}

Unlike spin-ordered materials with a vector order parameter, chirality arising in a non-magnetic material and without breaking time-reversal symmetry is unusual. Theoretically, the chiral state in TiSe$_2$ is proposed to originate from relative phase differences between the three CDW wave vectors present in this compound \,\cite{chiral_TiSe2_STM,jasper_EPL}. Having non-zero relative phases minimizes the local Coulomb interaction between charge density wave components. This is made possible in TiSe$_2$ by the fact that the modulations of the electronic density occur in distinct electronic orbitals for different CDW components\,\cite{jasper_EPL}. As a consequence, the chiral charge order will inevitably coincide with the emergence of orbital order.
 
Recent resonant elastic x-ray scattering (REXS) experiments revealed orbital order in ${\mathrm{TiSe}}_2$, with strong resonance at the Ti $K$ pre-edge\,\cite{TiSe2_PRR}. Here, we report REXS data collected at the Ti $K$-edge with different incident polarizations. The coupling between the helicity of right- or left-circular polarized x-rays to the screw axes in a crystal structure is widely used to explore the chirality in skyrmions\,\cite{skyrmion,nc_skyrmion}, polar vortices\,\cite{polar_vortex}, and chiral crystal structures\,\cite{Tanaka_2012,chiralXRD_PRL,chiralXRD_PRB,chiralRXS_Dy,E.N.Ovchinnikova}. We observe giant dichroism at reflections associated with the orbital order in the VdW material $1T$-TiSe$_2$, providing direct and unambiguous evidence for inversion symmetry breaking in its charge and orbital ordered state. Details of the experimental procedure can be found in the Methods section.

\begin{figure}[htbp]
\centering
\includegraphics[width=1\linewidth]{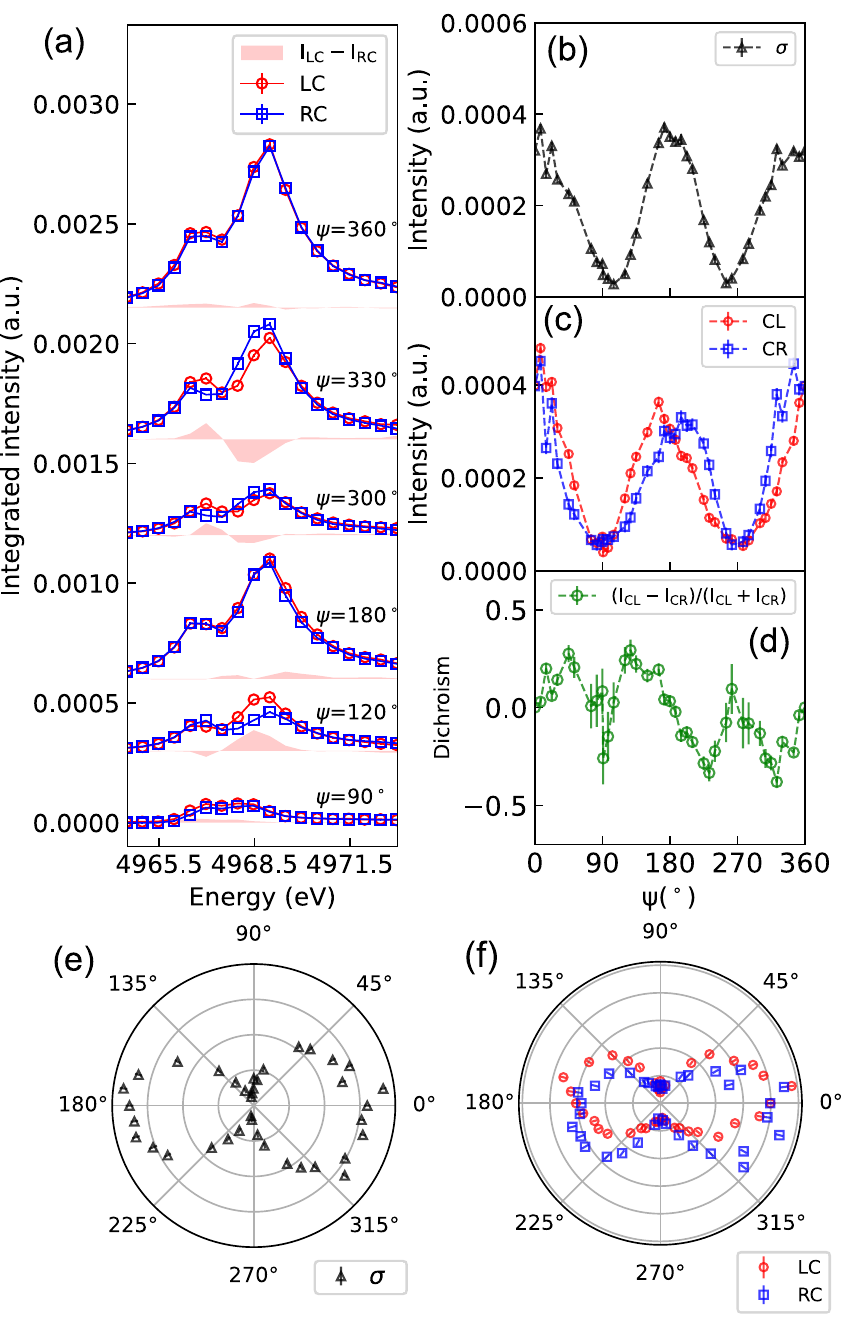}
\caption{(a) Energy dependence of the REXS intensity at the [0.5, 0, 2.5] reflection under LC and RC-polarized incident x-rays, for different azimuthal angles. The shaded areas indicate the difference between the LC and RC intensities. A vertical offset is added for different azimuthal angles for clarity. (b, c) Azimuthal dependence of the REXS intensity under $\sigma$-polarized incident x-rays (b), and using LC and RC polarization (c). (d) Azimuthal dependence of the circular dichroism, defined as ${\mathrm{(I_{LC} - I_{RC})/(I_{LC} + I_{RC})}}$. Error bars in (b-d) are estimated using the square root of the covariance value from the Gaussian fit to the peak profile, as well as measurement uncertainties. (e, f) Azimuthal dependence of the REXS intensity plotted in polar coordinates for $\sigma$-polarized incident x-rays (e), and LC and RC polarizations (f). The integrated intensity data were obtained by fitting the $\theta-2\theta$ scan at each incident photon energy with a Gaussian function (see Supplemental Material for details). 
\label{fig2}}
\end{figure}

The experimental geometry used in the REXS measurements is shown in Figure~\ref{fig1}(a). The crystal with ab-plane termination was attached to a wedge with an angle of 21.3$^\circ$. The scattering plane is vertical, defined by the wave vectors of the incident ({\bf k}$_i$) and outgoing ({\bf k}$_f$) x-rays. To assess the symmetries of the order parameter, three polarizations of incident x-rays were employed, as illustrated in Fig.~\ref{fig1}(b): $\sigma$ (orthogonal to the scattering plane), left-circular (LC), and right-circular (RC). The polarization of outgoing x-rays was not distinguished. X-ray absorption spectra (XAS) at the Ti $K$ edge were collected in total fluorescence yield (TFY) as shown in Fig.~\ref{fig1}(d). We observe no difference in the XAS spectra obtained with different polarizations (see Supplemental Material for further details). 

Our recent Ti K-absorption edge REXS study on 1T-TiSe$_2$ has revealed three qualitatively different groups of CDW reflections: conventional CDW peaks, reflections originating from orbital order and reflections originating from a mix of both\,\cite{TiSe2_PRR}. Here we focus on those reflections associated exclusively with the orbital order of ${\mathrm{TiSe}}_2$. The orbital order allows otherwise forbidden reflections to emerge at the pre-edge of the Ti $K$ absorption edge owing to hybridization between the Ti 3d orbitals and ligand Se 4p orbitals, as well as quadrupole-quadrupole transitions. Figure~\ref{fig1}(c) displays the intensity maps as a function of energy for the orbital order related peak at $\mathbf{Q} = \left[-0.5, 0.5, 2.5\right]$, under three different incident polarizations. The maps were obtained by performing $\theta-2\theta$ scans at each incident photon energy. The integrated intensity as a function of incident photon energy is shown in Fig.~\ref{fig1}(d). 

Owing to the high resolution of the current experiment, we were able to resolve a double-peak structure in the energy dependence of the scattering intensity and to go beyond previous REXS studies\,\cite{TiSe2_PRR}. The Ti atoms in ${\mathrm{TiSe}}_2$ are subject to an octahedral crystal field, which splits the 3$d$ energy levels into $t_{2g}$ and $e_g$ orbitals. The two peaks shown in the energy profile are related to scattering processes involving these two groups of orbitals\,\cite{TiSe2_L3L2XAS}. Intriguingly, the energy dependence of the scattering intensity shows clearly different responses to LC and RC-polarized incident x-rays, especially at 4967 and 4969\,eV, indicating a strong circular dichroism.

\begin{figure}[htbp]
\centering
\includegraphics[width=1\linewidth]{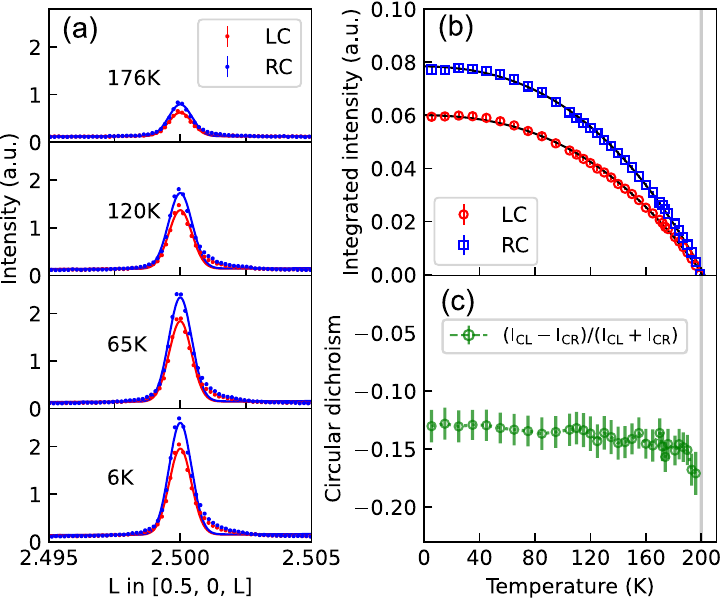}
\caption{(a) Selected $\theta-2\theta$ scans under LC and RC-polarized incident x-rays at $\psi = 210^\circ$ and $\mathbf{Q} = \left[0.5, 0, 2.5\right]$. The data have been normalized to the beam current, and error bars indicate the square root of the counts, which is smaller than the marker size. The solid lines are Gaussian fits of each scan. (b) Temperature evolution of the integrated intensity. Error bars represent the square root of the covariance value from the fit, which is smaller than the marker size. (c) Thermal evolution of the circular dichroism. The grey vertical line in (b) and (c) indicates the CDW onset temperature of 200\,K.
\label{fig3}}
\end{figure}

The energy dependence of the scattering intensity under both LC and RC incident x-rays is shown for several selected azimuthal angles $\psi$ in Fig.\,\ref{fig2}(a). The double-peak structure is robust and visible at all measured azimuthal angles. The spectra with LC- and RC-polarization overlap at $\psi = 90^\circ$, $180^\circ$ and $360^\circ$, while a clear difference at other $\psi$ angles points to an azimuthal evolution of the circular dichroism. Note that the measurement at $\psi = 270^\circ$ was omitted due to contamination by multiple scattering peaks. Interestingly, the sign of the observed dichroism is opposite for the two resonant energies, indicating a strong energy dependence as well as a relation of the dichroism to the orbital structure.

Focusing on the resonance at $E_i$ = 4968\,eV, a full scan of the azimuthal dependence of the scattering intensity under $\sigma$-, LC-, and RC-polarized x-rays is shown in Figs.\,\ref{fig2}(b-c). The rotation axis used here is $\mathbf{Q} = \left[0.5, 0, 2.5\right]$, which is equivalent to $\mathbf{Q} = \left[-0.5, 0.5, 2.5\right]$ under $C_3$ rotation symmetry. The azimuthal angle $\psi$ = 0$^\circ$ corresponds to the in-plane direction [1, 0, 0] pointing along the scattered beam. We find that the integrated intensity varies strongly with both the azimuthal angle $\psi$ and the polarization of the incident x-rays. The angular dependence in Fig.~\,\ref{fig2}(b) shows a nearly perfect mirror symmetry around $\psi$ = 180$^\circ$ for $\sigma$-polarized incident x-rays. The response to LC and RC polarized incident x-rays in Fig.~\,\ref{fig2}(c), however, is strikingly different, with the LC and RC curves approximately mirroring one another around $\psi$ = 180$^\circ$. This is visualized more clearly in polar coordinates, as shown in Fig.~\,\ref{fig2}(e-f).
The extracted circular dichroism as a function of azimuthal angle $\psi$ is shown in Fig.~\ref{fig2}(d), where the circular dichroism reaches values as high as $\sim$ 40$\%$.

\begin{figure}[htbp]
\centering
\includegraphics[width=1\linewidth]{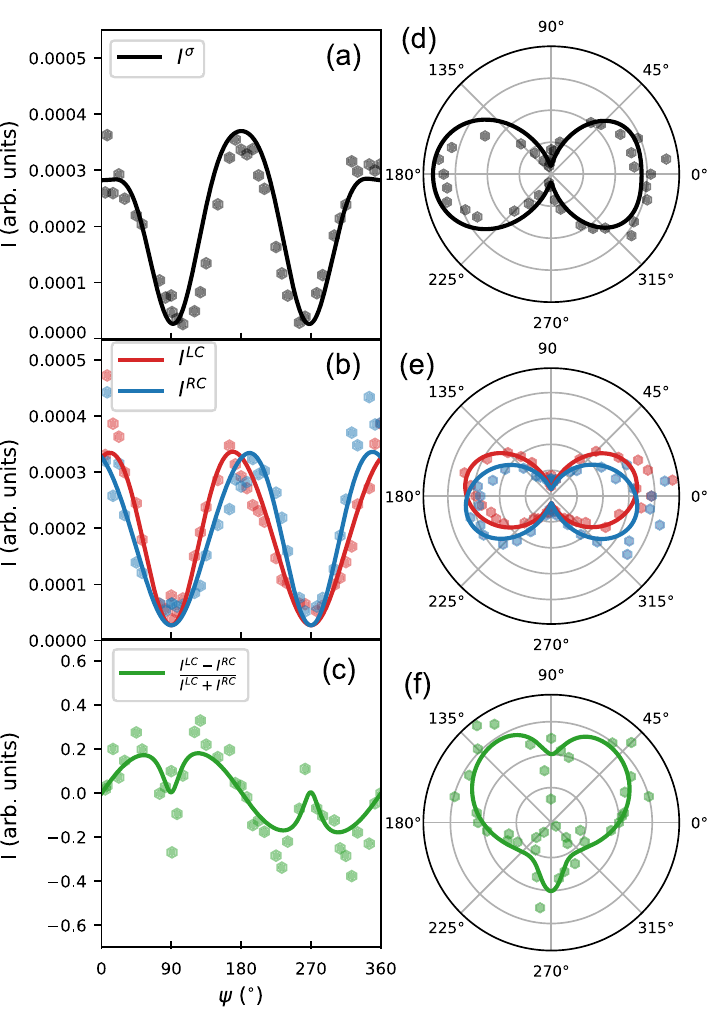}
\caption{Comparison of calculated (full line) and measured (shaded hexagons) azimuthal dependence of the REXS intensity under $\sigma$-polarized incident x-rays (a), and using LC and RC polarization (b). (c) Comparison of calculated (full line) and measured (shaded data) azimuthal dependence of the circular dichroism, defined as ${\mathrm{(I_{LC} - I_{RC})/(I_{LC} + I_{RC})}}$. (d, e, f) Same data as in panels (a, b, c) plotted in polar coordinates. All calculations employ the $C2$ structure from Ref.~\cite{TiSe2_PRR} with parameter values $A=0.01 a$, $\gamma=0.5$,
and $\delta=0.015$.
\label{fig4}}
\end{figure}

The onset temperature of the circular dichroism was determined by performing a temperature-dependent measurement at $\mathbf{Q} = \left[0.5, 0, 2.5\right]$ under LC- and RC-polarizations. For these measurements, the azimuthal angle $\psi$ was set to 210$^\circ$ and the incident energy $E_i$ was fixed at 4968\,eV. Some representative $\theta$-2$\theta$ scans at different temperatures are shown in Fig.\,\ref{fig3}(a) (for the full temperature range, see the Supplemental Material). The thermal evolution of the integrated intensity is displayed in Fig.\,\ref{fig3}(b). It can be seen to decrease smoothly with increasing temperature. The curves for both LC- and RC-polarization can be fitted using the empirical function $f(t) = C_1\{1-[(t+C_2)/(1+C_2)]^{C_3}$, where $t=T/T_{\mathrm{onset}}$ is the reduced temperature, and $C_1$, $C_2$ and $C_3$ are fitting parameters. This function has been used to fit the thermal evolution of the conventional CDW order parameter in 1T-TiSe$_2$\,\cite{fitting_Tdep_TiSe2}. The onset temperature of the giant dichroism is found to be $T_{\mathrm{onset}}$ $\simeq$ 200\,K, which coincides with the temperature where conventional CDW and orbital order first appear\,\cite{TiSe2_PRR}. The strength of the circular dichroism can be seen in Fig.\,\ref{fig3}(c) to be approximately constant across the entire temperature range, varying between 17$\%$ immediately below $T_{\mathrm{onset}}$ to approximately 13$\%$ at the lowest temperatures for $\psi$ at 210$^\circ$.

To track down the origin of the observed giant dichroism in 1T-TiSe$_2$, we perform numerical ab initio calculations of the scattering intensity (see Methods section and Supplemental Material for details). First, notice that in the space group $P\bar{3}c1$ expected for the non-chiral CDW phase of TiSe$_2$, the observed dichroism is prohibited by inversion symmetry. We confirm this by calculating the intensities of reflections at $\mathbf{Q}=\left[1, 0, 5\right]$ and $\mathbf{Q}=\left[\pm1, \mp1, 5\right]$ as a function of azimuthal angle (see Supplemental Material).

Next, we consider the inversion symmetry breaking induced by a non-zero relative phase shift between CDW components, as predicted theoretically in Refs.~\onlinecite{chiral_TiSe2_183K,TiSe2_PRR}. The model developed there considers three CDW components with equal amplitudes. The change in length of a single Ti-Se bond due to one of the CDW components is described by the displacements $A$ and $\gamma A$ for the Ti and Se atoms respectively. Typical values consistent with the structural data in Ref.~\onlinecite{1976prb} are $A=0.01 a$ and $\gamma=0.5$, with $a$ the in-plane lattice distance. In the proposed chiral phase, two of the CDW components obtain phase shifts (of opposite sign) compared to the third. This changes the displacements of Ti atoms along corresponding Ti-Se bonds to $(1\pm\delta)A$. The broken inversion symmetry of the resulting structure is controlled by $\delta$, and is described by the monoclinic space group $C2$ (see Supplemental Material for details). In this structure, the scattering vectors are labelled $\mathbf{Q}=\left[\pm1, \pm1, 5\right]$ and $\left[\pm2, 0, 5\right]$. By construction, the $\delta=0$ structure is equivalent to the $P\bar{3}c1$ CDW phase, and hence no dichroism is expected. In the calculations with $\delta=0$ in the $C2$ structure, however, the inversion symmetry is not analytically enforced, and numerical inaccuracies in the estimation of the electric field gradients result in minute amounts for the calculated dichroism. The maximal values are at least three orders of magnitude smaller than the expected $I_{\text{LC}}$ and $I_{\text{RC}}$, and are used to set error bars for the subsequent computational procedures.

Turning to the chiral structure with $\delta>0$, we find that inversion symmetry breaking drastically increases the polarization-averaged intensity as well as individual contributions in the different scattering channels. Most importantly, significant dichroism appears for a range of values of the inversion-breaking parameter $\delta$ and its angular dependence is strikingly similar to that observed in the experiment (see Fig.~\ref{fig4}). This agreement between experiment and theory is a direct indication that the model of the crystal structure correctly captures the qualitative symmetries of the chiral phase. Using the value of $\delta$ as a fitting parameter, we find that $\delta=0.015$ provides excellent quantitative agreement between experimental and computed data, as shown in Fig.~\ref{fig4}. Moreover, at this value of $\delta$, the calculated dichroism at the Se K-edge is very small (see Supplemental Material), in line with its apparent absence in earlier REXS studies~\cite{TiSe2_Seedge_prr}.

Our REXS measurements reveal a giant circular dichroism at specific diffraction peaks associated with the orbital order in ${\mathrm{TiSe}}_2$, which varies strongly with energy and azimuthal angle. Theoretical modeling of the scattering cross section for these peaks shows that they are forbidden reflections in the inversion-symmetric structure originally proposed for the CDW phase of $1T$-TiSe$_2$\,\cite{TiSe2_CDW}. The reflections instead become allowed in the inversion-broken electronic structure that is associated with its proposed chiral charge and orbital ordered phase\,\cite{chiral_TiSe2_STM,jasper_EPL,TiSe2_PRR}, and are found to cause circular dichroism. The characteristic azimuthal dependence of the observed dichroism is reproduced accurately in the model, with the amplitude of an inversion-breaking distortion $\delta$ as the only unconstrained fitting parameter.

It should be noted that previous REXS studies at the Se $K$-edge did not uncover chirality in TiSe$_2$\,\cite{TiSe2_Seedge_prr}. This is consistent with the chirality originating from orbital order which is associated primarily with the localized Ti 3d-electrons, rather than the delocalized Se p-electrons\,\cite{TiSe2_PRR}. Our observation of giant circular dichroism  establishes unambiguously that broken inversion symmetry is associated with orbital order in pristine TiSe$_2$ and opens the way to further explorations of the fundamental mechanisms linking charge density modulation, orbital occupation, and the spontaneous breakdown of symmetries or the emergence of chirality.\\

Y.\,Y.\,P.\ is grateful for financial support from the Ministry of Science and Technology of China (2021YFA1401903 and 2019YFA0308401) and the National Natural Science Foundation of China (Nos. 12374143 and 11974029). O.\,J.\ was supported by the Leibniz Association (Project No.\ J50/2018) through the Leibniz Competition. O.\,J.\ thanks Hiroki Ueda and Urs Staub for useful discussions, and Ulrike Nitzsche for the maintenance of the computational cluster at the IFW and technical support. W. X. thanks the Office of China Postdoc Council and the Helmholtz Association for funding for post-doctoral fellowship. We acknowledge DESY (Hamburg, Germany), a member of the Helmholtz Association HGF, for the provision of experimental facilities. REXS experiments were carried out at beamline P09 of PETRA III in mail-in mode due to traveling COVID restrictions. Beamtime was allocated for proposals I-20200132 and I-20211222.\\

\end{document}